
%
%
\magnification=1200
\tolerance=10000
\hsize 14.5truecm
\hoffset 1.25truecm
\font\cub=cmbx12
\baselineskip=22truept
\parindent=1.truecm
\def\etal{{et al.}\ }
\font\seventeenrm=cmr17
\centerline{\seventeenrm Testing the Frozen--Flow Approximation}
\vskip .5in
\centerline{{\bf Adrian L. Melott},$^1$ {\bf Francesco Lucchin},$^2$ {\bf
Sabino Matarrese},$^3$ and {\bf Lauro Moscardini}$^2$}
\vskip .5in
\item{1.} Department of Physics and Astronomy, University of Kansas, Lawrence,
Kansas 60645 USA
\vskip .25in
\item{2.} Dipartimento di Astronomia, Universit\`a di Padova, vicolo
dell'Osservatorio 5, I--35122 Padova, Italy
\vskip .25in
\item{3.} Dipartimento di Fisica ``G. Galilei", Universit\`a di Padova, via
Marzolo 8, I--35131 Padova, Italy
\vskip .50in
\baselineskip=15pt
\noindent {\bf Abstract:} We investigate the accuracy of the frozen--flow
approximation (FFA), recently proposed by Matarrese \etal (1992), for following
the nonlinear evolution of cosmological density fluctuations under
gravitational instability. We compare a number of statistics between results of
the FFA and nbody simulations, including those used by Melott, Pellman \&
Shandarin (1993) to test the Zel'dovich approximation. The FFA performs
reasonably well in a statistical sense, e.g. in reproducing the
counts--in--cell distribution, at small scales, but it does poorly in the
crosscorrelation with nbody which means it is generally not moving mass to the
right place, especially in models with high small--scale power.
\vfill\eject
\noindent {\bf 1. INTRODUCTION}

Gravitational instability is the dominant theory of how structure grew in the
universe. The primary tools for understanding this process have been linear
perturbation theory, the ``Zel'dovich approximation", and direct numerical
integration, usually called ``nbody simulations." For a review see Shandarin \&
Zel'dovich (1989).

Choosing the best approximation is for a given use is extremely important, as
approximations often form the basis for semianalytic arguments about galaxy
and/or large--scale structure formation, and are often used to provide initial
conditions or boundary conditions for nbody simulations.

The Zel'dovich (1970) approximation (ZA) was originally applied to so--called
``pancake" models, in which high--frequency modes are missing from the initial
mass density fluctuation spectrum. Beginning in Melott \etal (1983), evidence
began to appear that pancake--like structures might arise in models without
such damped initial conditions. More recently, ZA has been used to follow a
variety of models into the quasilinear regime. A variety of improvements on it
have been proposed.

Coles, Melott \& Shandarin (1993) (hereafter CMS) began systematic and
quantitative testing of some dynamical approximations, emphasizing the use of a
variety of initial conditions, and using crosscorrelation analysis to test
dynamics. Melott \etal (1993) (hereafter MPS)  did a detailed study of the
approximation CMS had determined to be the best, finding that it could be
considerably improved. The result was the truncated Zel'dovich approximation
(TZA), which is nothing more than the ZA with a specific filtering of initial
conditions. We used here the same nbody simulations and the same comparison
methods used in these two papers and in this work report results on the
frozen--flow approximation (FFA) as described in Matarrese \etal (1992)
(hereafter MLMS).

The plan of the paper is as follows. In section 2 we describe this
approximation, as well as the Zel'dovich approximation to which it is related.
In section 3 we define our main tool to compare the dynamics of different
approximations and nbody simulations: the crosscorrelation analysis. Section 4
presents the results of this analysis applied to FFA and TZA vs. nbody, as well
as some other statistics applied to the particle distributions. A final
discussion section concludes our paper.
\vskip .25in
\noindent {\bf 2. FROZEN--FLOW APPROXIMATION}

The standard Newtonian equations for the evolution of collisionless matter in
the universe can be rewritten in terms of suitably rescaled variables and in
comoving coordinates. In particular, it is sometimes convenient to use as time
variable the growth factor of linear perturbations, which in a flat, matter
dominated model, coincides with the expansion factor $a(t)=a_0(t/t_0)^{2/3}$ (a
subscript $0$ will be used for the ``initial time" $t_0$). The Euler equations
read
$$
{d {\bf u} \over d a} + {3 \over 2 a} {\bf u} = - {3 \over 2 a}
\nabla \varphi,
\eqno(1)
$$
where ${\bf u} \equiv d {\bf x}/ d a$ is a rescaled comoving peculiar velocity
field and the symbol ${d \over d a}$ stands for the total (convective)
derivative ${d \over d a}= {\partial \over \partial a} + {\bf u} \cdot \nabla$.
The continuity equation can be written in terms of the comoving matter density
$\eta({\bf x},t) \equiv \varrho({\bf x}, t) ~a^3(t) / \bar \varrho_0 a_0^3$
(where $\bar \varrho_0$ is the mean mass density at $t_0$)
$$
{d \eta \over d a} + \eta \nabla \cdot {\bf u} = 0,
\eqno(2)
$$
while the rescaled local gravitational potential $\varphi \equiv
(3t_0^2/2a_0^3) \phi({\bf x},t)$ is determined by local density inhomogeneities
$\delta({\bf x},t) \equiv \eta({\bf x},t) - 1$ through Poisson's equation
$$
\nabla^2 \varphi = {\delta \over a}.
\eqno(3)
$$
We restrict our analysis to irrotational flow.

The Zel'dovich approximation, in these variables, corresponds to the ansatz
${\bf u} = - \nabla \varphi$, as suggested by linear theory. In this case the
Euler and continuity equations decouple from Poisson's one, and the system
describes inertial motion of particles with initial velocity field impressed by
local gravity, as implied by the growing mode of linear perturbation theory:
${\bf u}_{ZA}({\bf x},\tau) = - \nabla_{\bf q} \varphi_0({\bf q})$, where ${\bf
q}$ is the initial (Lagrangian) position and $\tau \equiv a - a_0$. It follows
that particles move along straight trajectories
$$
{\bf x}({\bf q},\tau) = {\bf q} - \tau {\bf \nabla}_{\bf q}
\varphi_0({\bf q}).
\eqno(4)
$$

The frozen--flow approximation can be defined as the solution of the linearized
Euler equations, where in the r.h.s. the growing mode of the linear
gravitational force is assumed, ${\bf u}_{FFA}({\bf x},\tau) = {\bf u}_0({\bf
x}) = - \nabla_{\bf x} \varphi_0({\bf x})$, plus a negligible decaying mode. In
this approximation the peculiar velocity field ${\bf u}({\bf x},a)$ is {\it
frozen} at each point to its initial value, that is
$$
{\partial {\bf u} \over \partial \tau} =0,
\eqno(5)
$$
which is just the condition for steady flow. The above equation, together with
the continuity equation, define FFA. Particle trajectories in FFA are described
by the integral equation
$$
{\bf x}({\bf q},\tau) = {\bf q} - \int_0^\tau d \tau' \nabla_{\bf x}
\varphi_0[{\bf x}({\bf q},\tau')].
\eqno(6)
$$
Particles update their velocity at each infinitesimal step to the local value
of the linear velocity field, without any memory of their previous motion, i.e.
without {\it inertia}. Stream--lines are then frozen to their initial shape and
multi--stream regions cannot form. A particle moving according to FFA has zero
component of the velocity in a place where the same component of the initial
gravitational force is zero, it will then slow down its motion in that
direction while approaching that place. Unlike the Zel'dovich approximation,
these particles move along curved paths: once they come close to a pancake
configuration they curve their trajectories, moving almost parallel to it, and
trying to reach the position of the next filament. Again they cannot cross it,
so they modify their motion, while approaching it, to finally fall, for $\tau
\to \infty$, into the knots corresponding to the minima of the initial
gravitational potential. This type of dynamics implies an artificial thickening
of particles around pancakes, filaments and knots, which mimics the
gravitational clustering around these structures (though these configurations
do not necessarily occur in the right Eulerian locations, nor they necessarily
involve the right Lagrangian fluid elements). In assuming that the velocity
potential is linearly related at any time to the local value of the initial
gravitational potential, FFA disregards the non--linear effects caused by the
back--reaction of the evolving mass density on the peculiar velocity field
itself (via the non--linear evolution of the gravitational potential). This
implies that a number of physical processes such as merging of pancakes,
fragmentation and disruption of low--density bridges, are totally absent in the
FFA dynamics. Unlike the velocity field, the FFA density field is non--locally
determined by the initial fluctuations, via the continuity equation; this is
clearly shown by the following analytic expression
$$
1 + \delta_{FFA}({\bf x},\tau) = \exp \int_0^\tau d \tau'
\delta_+[{\bf x}({\bf q},\tau')]
\eqno(7)
$$
(where $\delta_+\equiv\delta_0/a_0$). Brainerd, Scherrer \& Villumsen (1993)
have recently shown that a similar formula also applies if one uses a different
approximation (called LEP: linear evolution of potential), consisting in
``freezing" the gravitational rather than the velocity potential [see also the
equivalent ``frozen--potential" approach by Bagla \& Padmanabhan (1993)]. This
approach shows many features in common to FFA, although multi--stream regions
do occur in this case.

Numerical implementation of FFA is straightforward (for a more technical
discussion, see MLMS) and involves small computing time: roughly speaking, FFA
consists of a multi--step Zel'dovich approximation, and very few steps are
required to follow the entire evolution. MLMS applied FFA to follow the
evolution of structures in the standard CDM model, and found that it gives a
fairly accurate representation of the density pattern from a resolution scale
of $\sim 500$ km s$^{-1}$, while the two--point correlation function fits quite
well the true non--linear result on even smaller scales. Further connections of
FFA and ZA can be found, based on the Hamilton--Jacobi approach to the
non--linear dynamics of collisionless matter. These, as well as other features
of FFA, will be discussed elsewhere.

\vskip .25in
\noindent {\bf 3. TESTING COMPARISONS}

We will use a group of nbody simulations, described in considerable detail
elsewhere (Melott \& Shandarin 1993, hereafter MS). MS used an ensemble of
simulations to get average values for a number of quantities and determine
which are the most stable statistics.  However, such ensembles are not
necessary to unearth systematic effects when everything else is held constant,
as verified by CMS. Here as in MPS, we use a subset of the ensemble: four
simulations with initial power--law density fluctuation power spectra
$$
P(k)=\mid\delta(k)\mid^2\propto k^n,
\eqno(8)
$$
with values $n=1,0,-1$, and $-2$. The case $n=-3$ is basically limited by
boundary conditions rather than the details of time evolution, and so is not
very interesting to follow. These Particle--Mesh nbody simulations have 128$^3$
particles on a 128$^3$ mesh, and are followed from very low amplitudes until
the clustering is so advanced that the absence of modes outside the box begins
to be a problem; this means an expansion factor of as much as 5000. All are
done in an $\Omega=1$ (Einstein--De Sitter) background to preserve
self--similarity as much as possible.

Stages are defined by the nonlinearity scale $k_{n\ell}$:
$$
a^2(t)\int^{k_{n\ell}}_0P(k)d^3 k = 1,
\eqno(9)
$$
where $P$ is the power in the initial conditions. This scale $k_{n\ell}$ is the
one which is, according to linear theory, going nonlinear. We study here
primary the stage $k_{n\ell}=8k_f$ where $k_f$ is the fundamental mode of the
box, but also crosschecked our results for consistency with $k_{n\ell}=4k_f$.

We will check a number of statistics for agreement between the models: the
power spectra, agreement of phases of Fourier components, the mass density
distribution, and its variance and skewness. We will examine the visual
appearance. But because we are testing dynamical approximations and not toy
models, we must also determine if mass has been moved to the right place;
statistical agreement is not enough.

Following CMS and MPS, we study the cross--correlation coefficient
$$
S={\langle\delta_1\delta_2\rangle\over \sigma_1\sigma_2},
\eqno(10)
$$
where $\sigma_i=\langle\delta_i^2\rangle^{1/2}$ and $\delta_1, \delta_2$ are
the pixellized density contrasts in the simulation and nbody distribution,
respectively. Of course $\mid S\mid\leq 1$, and $S=+1$ implies
$\delta_1=C\delta_2$ where $C$ is a constant.

This statistic has one overwhelming advantage over any other we can apply to
the mass distribution: as it approaches unity, {\it all} other statistics must
come into agreement. (Unless $S=1$ but $C\not= 1$; this is nearly impossible
here due to the way skewness grows with gravity.)

One possible problem with this approach is that an approximation might create
the right sort of condensations, but put them in slightly the wrong place. The
density peaks would miss each other and produce a small $S$. For this reason as
in CMS/MPS we also study smoothed fields. We smooth both $\delta_1$ and
$\delta_2$ by convolution with the same Gaussian $e^{-R^2/2R^2_G}$ and plot the
results as a function of $\sigma_2$ after smoothing. Thus, in the case above, a
large value of $S$ would appear with modest smoothing indicating good agreement
of the smoothed fields.

Brainerd \etal (1993) have criticized this approach. They studied evolution
based on LEP and on the ordinary ZA, and compared with an nbody simulation of
Cold Dark Matter. Both the nbody and the LEP produced small condensations and
ZA diffuse condensations in the same general region. Thus in some sense nbody
and LEP are more similar. Yet they found the unsmoothed crosscorrelation was
higher between nbody and ZA, and concluded that the value of crosscorrelation
was questionable for unsmoothed fields.

We disagree. In this test, LEP was penalized because it ``claimed" accuracy in
excess of what it had. Condensations had errors of position large compared to
their diameters. On the scale of the diameters the dynamics were wrong, as can
be seen by visually examining their plots. On the other hand, the more
conservative ZA represented its uncertainty by creating diffuse condensations,
which do include the analogous region in the nbody simulation, and produced a
stronger crosscorrelation. It is possible that LEP could crosscorrelate much
better than ZA or even TZA (see below) if examined with modest smoothing. The
pattern on medium scales appeared quite good.

The strategy in this paper, CMS, MPS, and near future work is to compare a
series of approximations in the same way. Thus we need to compare the
performance of FFA with some other approximation. The best that has emerged to
date is the Truncated Zel'dovich Approximation (TZA) as described by MPS.
Applying this approximation is simple, but results in a dramatic improvement
over ZA. One convolves the initial density field (still linear) with a Gaussian
$e^{-k^2/2k^2_G}$. The best choice value lies in the range $k_{n\ell}\leq
k_G\leq 1.5 k_{n\ell}$, depending on the spectrum (MPS). It may seem
paradoxical, but this smoothing of the initial conditions produces a less
smooth approximation in the nonlinear regime. It focuses pancakes where the
mass is, removing highly nonlinear modes which promote shell crossing and
diffuseness. As an example, the crosscorrelation for $n=-1,\sigma_2=2$ is
improved from 0.58 for ZA to 0.86 for TZA, with a similar improvement in visual
appearance (numbers quoted for $k_{n\ell}=4k_f$). We therefore will compare
statistics for nbody, FFA, and TZA; and the crosscorrelations of the latter two
with nbody will be compared.
\vskip .25in
\noindent {\bf 4. RESULTS}

In Figures 1--4 we can see the qualitative, visual effect of the approximation.
 These plots are greyscale renderings of the mass density in slices from the
nbody simulations, the FFA, and the TZA approximate solutions to the same
initial conditions.

Generally speaking, the patterns all look quite similar for $n = -2$, Figure 4.
 However, the FFA looks as if the flow were held back a bit. This tendency to
incomplete collapse is not serious here, but gets worse for more positive $n$.
For $n = +1$, Figure 1, the FFA appears to consist of many more very small
condensations than does the nbody simulation.  The patterns appear to have very
little in common.  Figures 1--4 (c), the TZA, seems to have more resemblance to
the nbody simulation, since the major condensations are of about the same size
in the right place.  TZA does seem to miss the small mass concentrations for
the more positive spectra, and it connects the larger ones by spurious
pancake--like bridges.  These bridges contain very very little mass, however.

The results of our crosscorrelation analysis are given in Figure 5. Basically,
two general results can be stated: (1) FFA performs better on an absolute scale
as $n$ decreases, and (2) TZA always performs better than FFA.  These
statements are in agreement with the visual impression. It is worth mentioning
at this time that FFA performs better than ordinary ZA for the spectra $n = 0$
and $n = +1$ in terms of crosscorrelation analysis.  Other workers (eg Brainerd
\etal 1993, Bagla \& Padmanabhan 1993) have compared their approximations to
ZA.  We have not done this for the reason that it is not difficult to
outperform ZA for positive spectra, since it performs so badly (see CMS).  So
far TZA appears to be the standard to beat.

The power spectra of the evolved distributions are plotted in Figure 6. The
results can be summarized easily. FFA underestimates small-$k$ power for all
initial conditions, and TZA is always quite good for small-$k$ power, agreeing
rather well up to about $0.6 k_{n\ell}$. Both approximations underestimate
large-$k$ power for all initial conditions, but FFA is always better than TZA.
This appears to be the reason that FFA does succeed in making many small dense
condensations. The normalization used here is one in which a Poisson
distribution with 128$^3$ particles would produce, on average, $P = 1$.

A clue to the low crosscorrelation is present in Figure 7, where we plot
$\langle\rm cos ~\theta\rangle$, where $\theta$ is the difference in phase
angle between a given Fourier component in the result of the nbody simulation
and that in its two approximate analogs. Of course 1 indicates perfect
agreement, and 0 total randomization.  For all spectra the phase agreement is
much better for TZA than for FFA.  MPS found in their experiments on the
effects of various windows for TZA that all windows produced similar spectra in
the results, but the phase agreement varied greatly, producing rather different
crosscorrelations.

The integrated mass density distribution function $F(>\rho)$ (with
clouds--in--cells binning of $128^3$ particles on our $64^3$ mesh) is plotted
in Figure 8, and confirms some of our suspicions.  In all cases except $n =
+1$, FFA reproduces a more peaked density distribution than TZA so that more
pixels reach high densities. Neither FFA nor TZA really produce a satisfactory
fit, except at moderate $\delta$ for $n = -2$, but the FFA appears to be better
overall. This is also the trend shown by the moment analysis of
counts--in--cells at the same scale (e.g. Lucchin \etal 1993, for a wider
introduction to this test); values for the variance, $\langle \delta^2\rangle$,
skewness $\langle \delta^3 \rangle$ (after shot--noise subtraction), for FFA,
TZA and nbody simulations, are reported in Table 1.
\vskip .25in
\noindent {\bf 5. DISCUSSION}

In summary, the FFA does not seem able to reproduce the dynamics, in the sense
of moving mass to the right places, at least for initial conditions with
substantial power on small scales.  The very simple TZA seems to work better in
this respect.  However, the FFA does succeed in producing small mass
condensations, the major point of failure of TZA; but it does not seem to put
them in the right place. Small--$k$ power grows too slowly, and large--$k$
power grows more correctly but with the wrong phases. This is probably because
of the way dynamics acts in FFA: particles move along the initial streamlines
to form ``first generation" pancakes, filaments and knots, but no merging of
these initial structures is allowed at all. Once the particles have come close
to these structures, all of the later clustering evolution consists of the slow
asymptotic fall of particles towards the wells of the initial gravitational
potential. No structures on larger scales will ever form. Thus, it is not
surprise that FFA provides better dynamical description in models with higher
large--scale power, where the first formed pancakes and filaments already
provide the large--scale structures. This is however an interesting feature, as
the most popular cosmological scenarios (such as cold or hot dark matter) have
low small--scale power. Also, FFA gives a better performance if evolved for
less expansion factors (as in $n=-2,-1$ models here), as otherwise the
large--scale pattern would significantly deviate both from nbody and linear
theory. We suspect that a similar trend would arise in the LEP approximation,
in spite of the different behaviour of particle trajectories near caustics
(e.g. Bagla \& Padmanabhan 1993). In order to improve the dynamical performance
of these approximations one would probably need either initial small--scale
smoothing or some account of the actual evolution of the velocity potential
beyond linear theory. Improvements of FFA along these lines will be discussed
elsewhere. The good statistical performance of FFA is not in contradiction with
the picture above. FFA seems to produce enough (or even too much) small--scale
structure: cell count statistics on a given scale will generally show the
correct trend, being only sensitive to the number of cells with the right
occupation frequency, not to their exact location.

Future work will involve tests of the adhesion approximation (Shandarin \&
Zeldovich 1989), the Linear Evolution of Potential approximation (Brainerd
\etal 1993; see also Bagla \& Padmanabhan 1993), and higher order
Zeldovich-like solutions (Buchert 1993 and references therein). It will be
interesting to see whether any of the proposed improvements really go beyond
the accuracy of the Zeldovich approximation, with truncation of modes that are
too nonlinear to follow. So far, that 1970 proposal, with Gaussian filtering of
modes that will act as noise, appears remarkably robust.
\vskip .25in
\noindent {\bf ACKNOWLEDGEMENTS}

ALM gratefully acknowledges financial support from NSF grants AST--9021414,
OSR--9255223, NASA Grant NAGW--2923, and computing support from the National
Center for Supercomputing Applications, Urbana, Illinois, USA. FL, SM and LM
acknowledge the Italian MURST for partial financial support, and the CINECA
Computer Center (Bologna, Italy) for the use of computing facilities.
\vfill\eject
\noindent {\bf REFERENCES}

\def\ref{\par\noindent\hangindent\parindent\hangafter1}
\ref
Bagla J.S., Padmanabhan T., 1993, MNRAS, in press

\ref
Brainerd T.G., Scherrer R.J., Villumsen J.V., 1993, ApJ, in press

\ref
Buchert T., 1993, MNRAS, submitted

\ref
Coles P., Melott A.L., Shandarin S.F., 1993, MNRAS, 260, 765 (CMS)

\ref
Lucchin F., Matarrese S., Melott A.L., Moscardini L., 1993,
ApJ, submitted

\ref
Matarrese S., Lucchin F., Moscardini L., Saez D., 1992, MNRAS, 259,
437 (MLMS)

\ref
Melott A.L., Einasto J., Saar E., Suisalu I., Klypin A., Shandarin
S.F., 1983, Phys. Rev. Lett., 51, 935

\ref
Melott A.L., Pellman T.F, Shandarin S.F., 1993, MNRAS, submitted (MPS)

\ref
Melott A.L., Shandarin S.F., 1993, ApJ, 410, 469 (MS)

\ref
Shandarin S.F., Zel'dovich Ya.B., 1989, Rev. Mod. Phys., 61, 185

\ref
Zel'dovich Ya.B., 1970, A\&A., 5, 84
\vfill\eject

\centerline{\bf Figure captions}
\bigskip
\bigskip
\noindent
{\bf Figure 1}. A greyscale plot of thin (L/128) slices of the simulation cube,
and the approximations to it for index $n=+1$ initial conditions at the stage
$k_{n\ell}=8k_f$. (a) the nbody simulation (b) the frozen--flow approximation
FFA (c) the Gaussian--truncated TZA model.
\medskip
\noindent
{\bf Figure 2}. As in Figure 1, but for $n=0$ initial conditions.
\medskip
\noindent
{\bf Figure 3}. As in Figure 1, but for $n=-1$ initial conditions.
\medskip
\noindent
{\bf Figure 4}. As in Figure 1, but for $n=-2$ initial conditions.
\medskip
\noindent
{\bf Figure 5}. A plot of the crosscorrelation $S$ as defined in the text
between the density field generated by the full nbody simulations and the
approximations versus the rms density fluctuation in the simulation. Both are
smoothed by the same Gaussian window and refer to $k_{n\ell}=8k_f$. Solid line:
the frozen--flow approximation, FFA. Dashed line: the Gaussian--truncated TZA
model.
\medskip
\noindent
{\bf Figure 6}. Power spectra at $k_{n\ell}=8k_f$ for the nbody simulations
(dotted--dashed line), for FFA (solid line) and for Gaussian TZA (dashed line).
\medskip
\noindent
{\bf Figure 7}. The average effective phase angle error, measured by $\langle
\rm cos ~\theta \rangle$ as described in the text, at the stage
$k_{n\ell}=8k_f$. Solid line: the frozen--flow approximation, FFA. Dashed line:
the Gaussian--truncated TZA model.
\medskip
\noindent
{\bf Figure 8}. The cumulative mass density distribution $F(>\rho)$ in terms of
the number of cells with given density $\rho$, in units of the mean density,
with clouds--in--cells binning of $128^3$ particles on our $64^3$ mesh.

\vfill\eject

\baselineskip=15truept

\bigskip\bigskip
\centerline {{\bf Table 1.} Moments of the density distribution.}
\bigskip
\bigskip
\centerline {\bf {variance}}
\bigskip
{\settabs 4\columns
\+ & nbody & FFA & TZA \cr
\+ & & & \cr
\+ $n=+1$   & ~~9.6 & ~1.7 & ~1.8 \cr
\+ $n=\hskip2.64mm 0$   & ~~9.3 & ~3.3 & ~1.9 \cr
\+ $n=-1$   & ~~8.3 & ~4.1 & ~1.5 \cr
\+ $n=-2$   & ~~8.5 & ~3.2 & ~2.6 \cr}
\bigskip
\bigskip
\centerline {\bf {skewness}}
\bigskip
{\settabs 4\columns
\+ & nbody & FFA & TZA \cr
\+ & & & \cr
\+ $n=+1$   & ~~240 & ~~7 & ~10 \cr
\+ $n=\hskip2.64mm 0$  & ~~279 & ~28 & ~11 \cr
\+ $n=-1$   & ~~298 & ~47 & ~11 \cr
\+ $n=-2$   & ~~455 & ~38 & ~26 \cr}
\bye